\documentclass[aps,prb,twocolumn,superscriptaddress,amssymb,floatfix,showpacs]{revtex4}
\usepackage{graphicx}
\usepackage{epstopdf}
\usepackage{soul}
\usepackage{ulem}
\usepackage[dvipsnames,usenames]{color}

\def\uSR{$\mu$SR }
\def\ZrNiGa{ZrNi$_2$Ga}

\begin{document}

\title{Probing the superconducting state of the Heusler superconductor: \ZrNiGa}


\author{A. D. Hillier}
\affiliation{ISIS facility, STFC Rutherford Appleton Laboratory, Harwell Science and Innovation Campus, Oxfordshire, OX11 0QX, UK}
\author{N. Parzyk}
\author{D. M$^{c}$K. Paul}
\affiliation{Physics Department, University of Warwick, Coventry, CV4 7AL, United Kingdom}


\date{\today}

\begin{abstract}
Using both muon spin rotation and muon spin relaxation, the superconducting ground state of the Heusler superconductor {\ZrNiGa} has been studied. The temperature dependence of the magnetic penetration depth of {\ZrNiGa} is consistent with a single isotropic gap s-wave BCS superconductor. The gap energy is $\Delta(0)$=0.44(1)~meV and the magnetic penetration depth, $\lambda(0)$, is 310(5)~nm. Furthermore, we show evidence of a possible cross-over from an square flux line lattice to a hexagonal lattice at low temperatures. No evidence of time reversal symmetry breaking has been observed as might be expected for a half metal superconductor.

\end{abstract}


\maketitle

\begin{figure}
\includegraphics[width=5 cm,clip=true, trim= 80 -30 0 121]{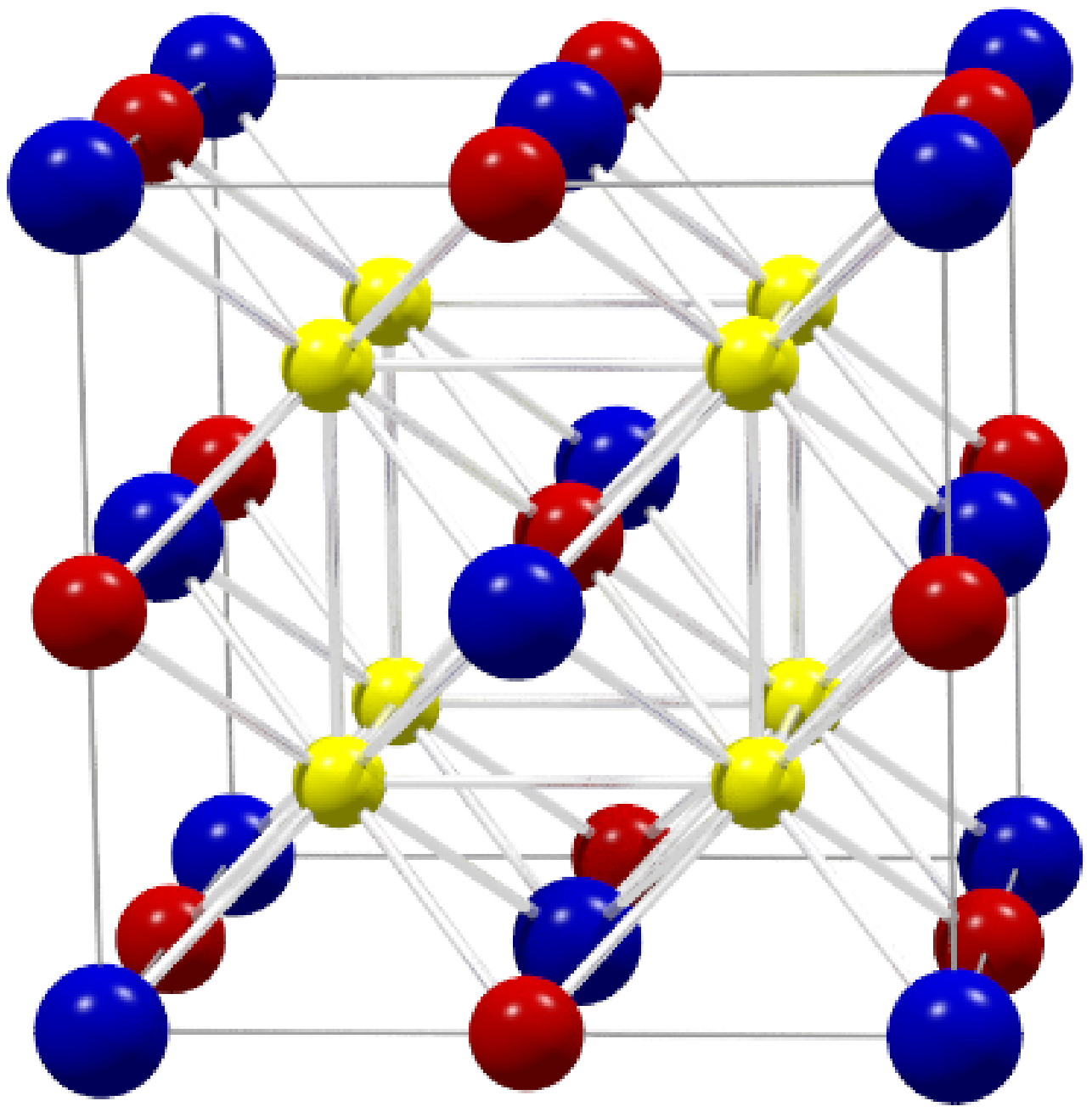}
\caption{\label{Fig:Struct} (color online) The  crystal structure of \ZrNiGa. The blue spheres (largest) are Zr, red spheres (medium) are Ga and the yellow spheres (smallest) are Ni. }
\end{figure}

Heusler compounds have attracted a great deal of interest since their discovery over 100 years ago\cite{Heusler1903a,Heusler1903b}. The search for appropriate new materials for spintronics applications suggested that the Heusler compounds might be good candidates for many applications. This, in part, was due to their extremely tunable electronic structure. There are now over 1000 known Heusler compounds, with properties which include shape-memory materials, Kondo systems, half-metallic ferromagnets, thermoelectric materials, compensated ferrimagnets and indeed even superconductors and topological insulators/superconductors\cite{Graf11}. These Heusler materials are ternary alloys, which can be broken down into two classes: the 2:1:1 (X$_2$YZ) or the 1:1:1 (XYZ half-Heusler) stoichiometry. The X$_2$YZ Heusler compounds crystallise with four interpenetrating fcc sublattices (see Fig. \ref{Fig:Struct}). 

To date, there have only been a few Heusler superconductors found: namely YPd$_2$Sn\cite{Wernick83}, NbNi$_2$X (X=Al, Ga, Sn)\cite{Wernick83,Waki85} and \ZrNiGa\cite{Winterlik08} with transition temperatures of 4.9~K, 2.15~K, 1.54~K, 3.4~K and 2.8~K respectively. Recently, there has also been the discovery of LaPtBi and YPtBi, which is a non-centrosymmetric superconductor with a low carrier density. The exchange splitting of the d-electron states and the close proximity to ferromagnetic order makes these Heusler superconductors an attractive prospect for unconventional superconductivity\cite{Goll08,Butch11,Chung11,Eschrig03}. With electrons of one-spin direction having a finite density of states at the Fermi level then this would make these Heusler superconductors a system in which one might expect non-unitary superconductivity. Evidence for non-unitary superconductivity would be exhibited by the breaking of time-reversal symmetry (TRS). In the broken TRS state  spontaneous fields appear. Although, there are a few but growing number of superconductors which break TRS, of those there are a small number which are non-unitary, namely LaNiC$_2$\cite{Hillier09}, PrOs$_4$Sb$_{12}$\cite{Aoki03,Shu11}, UGe$_2$\cite{Saxena00} and URhGe\cite{Aoki01}. One of the most direct ways of detecting an unconventional superconducting state is muon spin relaxation and rotation ($\mu$SR), as it can unambiguously establish broken time reversal symmetry by detecting the presence of a magnetic field in the sample when zero magnetic field is applied. Also, by using muon spin rotation the symmetry of the superconducting gap, structure of the flux line lattice as well as the magnetic penetration depth, $\lambda(0)$ can be determined. Therefore, in order to fully investigate the superconducting ground state of {\ZrNiGa} we have employed the $\mu$SR technique.

{\ZrNiGa} is a superconductor with a transition temperature of 2.8~K and has a full Heusler structure (see Fig. \ref{Fig:Struct}). Specific heat and magnetization measurements have shown that {\ZrNiGa} is a bulk type-II superconductor which is weakly coupled\cite{Winterlik08}. In the normal state, despite the enhanced Sommerfeld constant and the Van Hove singularity no magnetic order is observed and therefore {\ZrNiGa} can best be described as an enhanced Pauli magnet.  

In this communication we report $\mu$SR results for the superconductor {\ZrNiGa} showing that TRS is not broken on entering the superconducting state. We have also investigated the superconducting ground state and have determined the magnetic penetration depth, $\lambda(0)$,along with examining the temperature dependence of the magnetic penetration depth. Using these results we conclude that the {\ZrNiGa} is an s-wave superconductor and we find evidence for a possible square to hexagonal transition in the flux line lattice.

The sample was prepared by melting together stoichiometric amounts of the constituent elements in a water-cooled arc furnace under an atmosphere of argon. The resulting ingot was flipped several times to ensure sample homogeneity. The sample was then wrapped in Ta foil, sealed in an evacuated quartz tube and annealed at ${800^{o}}$C for 2 weeks then quenched into iced water.  The $\mu$SR experiments were carried out using the MuSR spectrometer in both geometries. At the ISIS facility, a pulse of muons is produced every 20 ms and has a FWHM of $\sim$70 ns. These muons are implanted into the sample and decay with a half-life of 2.2~$\mu$s into a positron which is emitted preferentially in the direction of the muon spin axis and two neutrinos. These positrons are detected and time stamped in the detectors which are positioned either before, F, and after, B, the sample for longitudinal (relaxation) experiments. Using these counts the asymmetry in the position emission can be determined and, therefore, the muon polarisation is measured as a function of time. For the transverse field experiments, the magnetic field was applied perpendicular to the initial muon spin direction and momentum.
The sample was powdered and mounted onto a 99.995+~$\%$ pure silver plate, with a 30mm diameter. Any muons stopped in silver give a time independent background for longitudinal (relaxation) experiments. The sample holder and sample were mounted onto a TBT dilution fridge with a temperature range of 0.045-4~K. The sample was cooled to base temperature in zero field and the $\mu$SR spectra were collected upon warming the sample while still in zero field. The stray fields at the sample position are cancelled to within 1~$\mu$T by a flux-gate magnetometer and an active compensation system controlling three pairs of correction coils. The (TF-$\mu$SR) experiment was conducted with applied fields between 5~mT and 60~mT, which ensured the sample was in the mixed state. Each field was applied above the superconducting transition before cooling. The MuSR spectrometer comprises 64 detectors. In software, each detector is normalised for the muon decay and reduced into a real and imaginary component. 

\begin{figure}
\includegraphics[width=8.0cm]{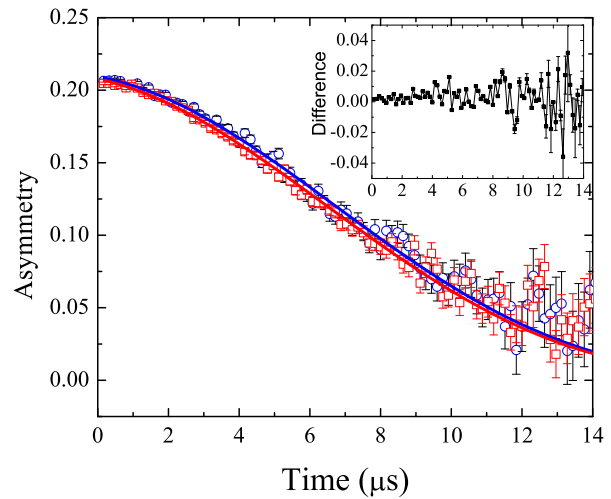}%
\caption{\label{fig:SpectraL} (color online) The zero field $\mu$SR spectra for{ \ZrNiGa}.  The red symbols are the data collected at 0.1~K and the blue symbols are the data collected at 3.0~K. The inset shows the difference of the spectra.}
\end{figure}

Considering the longitudinal $\mu$SR data. The absence of a  precession signal in the $\mu$SR spectra at all temperatures confirms that there are no spontaneous coherent internal magnetic fields associated with long range magnetic order in {\ZrNiGa} at any temperature. In the absence of atomic moments muon spin relaxation is expected to arise entirely from the local fields associated with the nuclear moments. These nuclear spins are static, on the time scale of the muon precession, and randomly orientated. The depolarisation function, $G_z(t)$, can be described by the Kubo-Toyabe function\cite{Hayano79}. In Fig. \ref{fig:SpectraL} we can see the data has the characteristic shape of the Kubo-Toyabe function, with a depolarisation rate of 0.093(1)~$\mu s^{-1}$ and 0.094(1)~$\mu s^{-1}$ for low and high temperatures respectively. More importantly, this doesn't change as a function of temperature. This indicates that time reversal symmetry is preserved, i.e. not broken, or at least any symmetry breaking field is not observable by $\mu$SR. 

\begin{figure}[tbh]

\includegraphics[width=8.0cm]{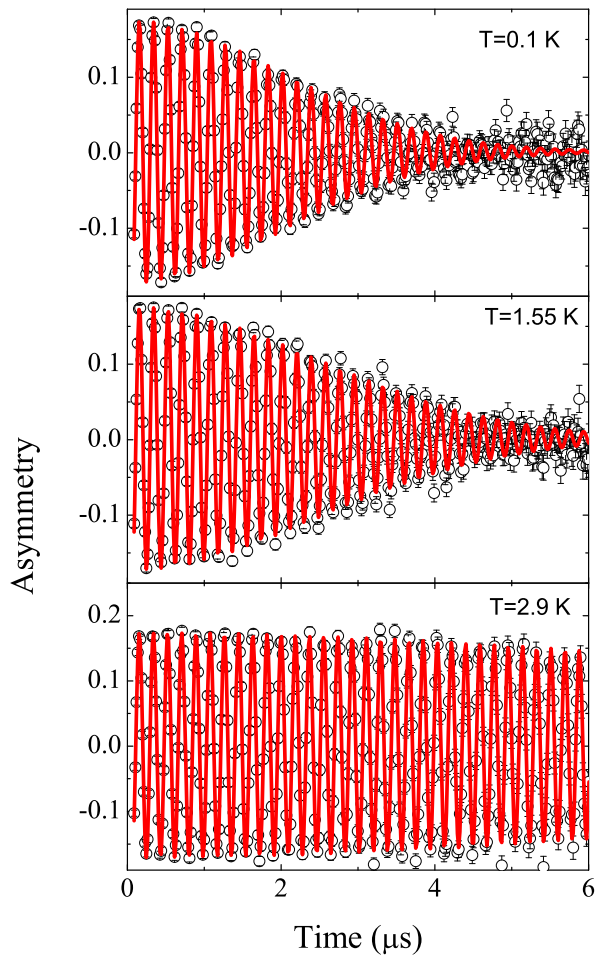}
\caption{\label{fig:Spectra}(Color online) Typical muon asymmetry spectra in {\ZrNiGa} taken in a transverse field of 40 mT at 0.1 (upper), 1.55 (middle) and 2.9 K (lower). The line is a fit to the data using Eqn. \ref{eqn:fitfunctf}. For clarity, only one of the two virtual detectors have been shown.}
\end{figure}

In an applied transverse magnetic field the experimental data were fitted with a sinusoidal oscillating function with a Gaussian relaxation component:

\begin{equation}
G_x(t) = \displaystyle\sum_{i=1}^2 A_i exp(-\frac{\sigma_i ^2t^2}{2})cos(2\pi \nu_i t + \varphi)
\label{eqn:fitfunctf}
\end{equation}

\noindent where the index \emph{i} denotes the contribution from the superconducting phase or the background, respectively, $A_i$ is the initial asymmetry, $\sigma_i$ is the Gaussian relaxation rate, $\nu_i$ is the muon spin precession frequency and $\varphi$ is the phase offset. The background term comes from those muons which were implanted into the silver sample holder and this oscillating term has no depolarisation, i.e. $\sigma_2$=0.0~$\mu s^{-1}$, as silver has a negligible nuclear moment. The contribution to the signal from the silver sample holder was negligible and set to zero in the data analysis. Fig. \ref{fig:Spectra} shows typical spectra for {\ZrNiGa} with an applied field of 40~mT at 0.1, 1.55 and 2.9~K after being field cooled through $T_{C}$. 

\begin{figure}[tbh]
\includegraphics[width=8.0cm]{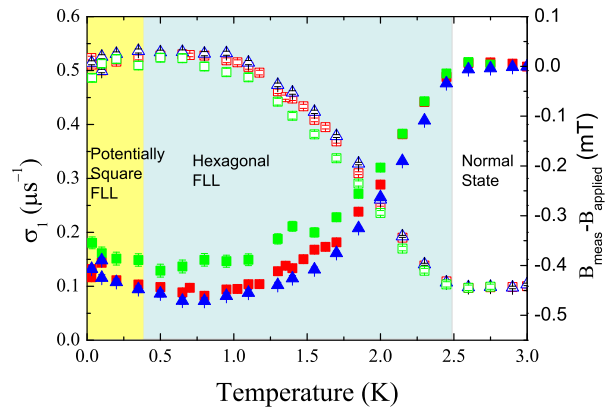}
\caption{\label{fig:SvT}(Color online) The temperature dependence of $\sigma$(hollow symbols) and $\Delta B$ (filled symbols) for fields of 30 (blue), 40(red) and 60~mT (green).}
\end{figure}

Fig. \ref{fig:SvT} shows the temperature dependence of the muon depolarisation rate $\sigma$. This shows an increase as the temperature is reduced through T$_C$. As the temperature is reduced still further the $\sigma$ levels then starts to slightly reduce. This effect is apparent in all fields and is only slightly temperature dependent. $\sigma$ is directly related to $\lambda$ by the relation,

\begin{equation}
\sigma_{sc}[\mu s^{-1}]=A(1-b)(1+1.21(1-\sqrt{b})^3)\lambda^{-2}[nm]
\label{eqn:Brandt}
\end{equation}  

\noindent where b=B/B$_{C2}$ the ratio of applied field to the upper critical field and A is a prefactor relating to the structure of the flux line lattice (A=$4.83\times 10^4$ and $5.07\times10^4$ for a hexagonal and square lattice respectively)\cite{Brandt03}. The superfluid density is directly related to $\lambda$ therefore it is difficult to understand a reduction in superfluid density at the temperature is reduced. Now, considering the mean field  value coming as observed from the \uSR spectra, then at the same temperature in which the decrease in $\sigma_1$ is observed then an decrease in the field shifts, $\Delta B$, is also observed\cite{Brandt88,Sidorenko90}. This change in mean field  value could indicate a structural change in the flux line lattice, which could be square (low temp, T \textless~0.4~K) to hexagonal (high temperature, T \textgreater~0.4~K) giving the reduction of $\sigma$ and $\Delta B$. Now considering the values for $\sigma$ at the peak and for T=0 then magnetic penetration depth can be determined, now assuming a square flux line lattice at low temperatures and hexagonal flux line lattice at higher temperatures, $\lambda$ are equal, within error, at 310(5) nm.

\begin{figure}[tbh]

\includegraphics[width=8.0cm]{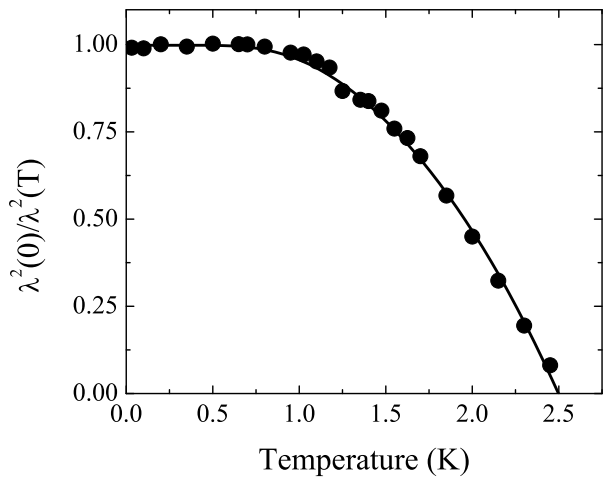}
\caption{\label{fig:TdepL} The temperature dependence of the $\lambda^2(0)/\lambda^2(T)$. The line is a fit to the data using an isotropic s-wave model (see text for details).}
\end{figure}

Now we present a detailed analysis of the $\sigma_{sc}(T)$. The $\sigma_1$ value measured whilst in the superconducting state, $\sigma_{sc}$ is a convolution of both the flux line lattice and the nuclear moments. 
The contribution from the nuclear moments has been determined at a temperature just above T$_C$ and has been assumed to be constant. This assumption is justified by the longitudinal zero field \uSR experiment discussed earlier. Now using eqn. \ref{eqn:Brandt} the temperature dependence of $\lambda$ can be determined. Here we have assumed that we have a square lattice below T \textless~0.4~K and hexagonal above. Clearly in Fig. \ref{fig:TdepL} $\lambda^{-2}(T)/\lambda^{-2}(0)$ shows a flat temperature dependence after correcting for the change in symmetry of the fluxline lattice at low temperatures (T \textless~0.4~K) and hence supports the possibility that there may be a structural change in the flux line lattice. The superconducting gap can be modelled by 

\begin{equation}
\frac{\lambda^{-2}(T)}{\lambda^{-2}(0)}=
1+2\int_{\Delta(T)}^{\infty} \left(\frac{\partial f}{\partial E}\right) 
\frac{E}{\sqrt{E^2-\Delta(T)^2}} dE
\end{equation}

\noindent where $f=\left[1+\exp\left(E/k_BT\right)\right]^{-1}$ which is the Fermi function\cite{Tinkham}. The temperature dependence of the gap was approximated by\cite{Carrington03}
\begin{equation}
 \Delta(T)=\Delta(0){\rm tanh}[1.82(1.018(T_c/T-1))^{0.51}].
\end{equation} 

As can be seen from Fig. \ref{fig:TdepL} the temperature dependence of $\lambda$ is very well described by an isotropic s-wave model giving $\Delta(0)$=0.44(1)~meV.

Now considering the implication of these results, {\ZrNiGa} appears to be a conventional isotropic s-wave superconductor, however, being a Heusler superconductor the half filled band at the Fermi surface should lead to non-unitary superconductivity and hence time-reversal symmetry breaking and  non isotropic s-wave temperature dependence\cite{Chung11,Eschrig03}. Neither of these characteristics are experimentally observed for this compound. Therefore, we can only conclude that this half filled band at the Fermi surface is not the band responsible for the superconductivity or that this band is not half filled.

In conclusion, $\mu$SR experiments have been carried out on {\ZrNiGa}. The zero field measurements do not show any spontaneous fields appearing below the superconducting transition temperature (T$_C$=2.8~K). This provides convincing evidence that time reversal symmetry is not broken in the superconducting state of this material. Indeed, we have probed the superconducting gap and found that the temperature dependence of the superconducting gap can be described by an isotropic s-wave gap with a T=0~K amplitude of $\Delta(0)$=0.44~meV. This implies that the half-filled bands at the Fermi surface are not those responsible for the superconductivity. However, there is an unusual temperature dependence in $\sigma$, which shows a small decrease in value, which coincides with an increase in the mean field value. We conclude that this is evidence of a structural change in the flux line lattice may result from a square to hexagonal transition. Clearly, single crystal studies using both muons and small-angle neutron diffraction are highly desirable.

\begin{acknowledgements}
This work was supported by the Engineering and Physical Sciences Research Council (EPSRC) and the Science and Technology Facilities Council (STFC) of the UK. 
\end{acknowledgements}


\bibliography{ZrNi2Ga_prbrc2}
\appendix
\end{document}